\documentclass[namedreferences]{SolarPhysics}
\usepackage[optionalrh]{spr-sola-addons} 
\usepackage{graphicx}        
\usepackage{color}           
\usepackage{url}             

\newcommand{\beq}{\begin{equation}}
\newcommand{\eeq}{\end{equation}}
\newcommand{\beqa}{\begin{eqnarray}}
\newcommand{\eeqa}{\end{eqnarray}}

\begin{document}


\begin{opening}

\title{On Thermal-Pulse-Driven Plasma Flows in Coronal Funnels as 
Observed by {\it Hinode}/{\it EUV} {\it Imaging} {\it Spectrometer} (EIS)}

\author{A.K.~\surname{Srivastava}$^{1}$\sep
    P.~\surname{Konkol}$^{2}$\sep
    K.~\surname{Murawski}$^{2}$\sep
    B.N.~\surname{Dwivedi}$^{1}$\sep
    A.~\surname{Mohan}$^{1}$}
    
\runningauthor{A.K.~Srivastava {\it et al.}}
\runningtitle{Thermal Pulse Driven Plasma Flows}
\institute{$^{1}$ Department of Physics, Indian Institute of Technology (Banaras Hindu University), Varanasi-221005, India.\\
                     email: \url{asrivastava.app@iitbhu.ac.in}\\
             $^{2}$Group of Astrophysics,
             UMCS, ul. Radziszewskiego 10, 20-031 Lublin, Poland.\\
email: \url{kmurawski@umcs.lublin.pl}
             }

\begin{abstract}
Using one-arcsecond-slit scan observations from the {\it Hinode}/{\it EUV} {\it Imaging} {\it Spectrometer} (EIS) on 05 February 2007, we find the plasma outflows in the open and expanding coronal funnels at the eastern boundary of AR 10940. 
The Doppler velocity map of Fe \textsc{xii} 195.120 \AA~shows that the diffuse close-loop system to be mostly red-shifted. 
The open arches (funnels) at the eastern boundary of AR exhibit blue-shifts with a maximum speed of about 10\,--\,15 km s$^{-1}$. This implies outflowing plasma through these magnetic structures. In support of these observations, we perform a 2D numerical
simulation of the expanding coronal funnels by solving the set of ideal MHD equations 
in appropriate VAL-III C initial temperature conditions using the FLASH code.
We implement a rarefied and hotter region at the footpoint of the model funnel, which results 
in the evolution of slow plasma perturbations propagating
outward in the form of plasma flows. 
We conclude that the heating, which may result from magnetic reconnection, 
can trigger the observed plasma outflows in such coronal funnels.
This can transport mass into the higher corona, giving rise to the formation of the nascent solar wind.
\end{abstract}
\end{opening}

\section{Introduction}

The solar wind is the supersonic outflow of fully ionized gas from the solar corona streaming through 
the magnetic field lines.
It is well established that the polar coronal hole is the source of fast
solar wind (Hassler {\it et al.,} 1999; Wilhelm {\it et al.,} 2000; Tu {\it et al.,}~2005)
while the slow solar wind originates from the boundary of active regions
and along the streamers in the equatorial corona (Habbal {\it et al.,} 1997; Sakao, 2007).
Hassler {\it et al.} (1999) have found that the plasma can be supplied
from the chromospheric heights in the network boundaries to the
solar wind in polar coronal holes. More precise estimate of the formation of nascent
wind in the coronal funnels between 5\,--\,20 Mm in coronal holes has been carried out by Tu {\it et al.} (2005).
The bases of the polar coronal holes are mostly the sources of
fast  solar wind. It has also been suggested that plasma outflows observed at the edges of active regions are the 
source of the slow solar wind. Active-region arches that can extend outward as the rays in the outer corona
may channel it (Slemzin {\it et al.,} 2013). In addition to the large-scale origin, it has also been found that the
small-scale outflows at the coronal hole boundaries (CHBs) can serve as the source of the slow solar wind
(Subramanian, Madjarska, and Doyle, 2010). Recently, Yang {\it et al.} (2013) have reported a numerical model to describe the
process of magnetic reconnection between moving magnetic features (MMFs) and the pre-existing ambient magnetic field
that drives anemone jet with inverted y-shape base and associated plasma blobs. They have found that an increase in
the thermal pressure at the base of the jet is also driven by the reconnection, which induces a train of slow-mode
shocks propagating upward resulting in plasma upflows. Their findings contribute to the formation of  
jets and small-scale flows in the quiet-Sun corona where MMFs undergo into the 
low atmospheric reconnection.

Two outstanding issues, however, remain unsettled: i)
what are the drivers of these winds in the outer corona? and ii) what are the source regions
and what drivers enable the mass supply to the lower solar atmosphere (Chromosphere-TR, and inner corona)?. 
There exist several
studies advocating the role of Alfv\'en waves as a possible answer to the first question. The ion-cyclotron waves
at kinetic scales are visualized as one of the possible candidates to provide momentum and heat the outer
coronal winds both in theory and observations ({\it e.g.} Ofman and Davila, 1995;
Tu and Marsch, 1997; Suzuki and Inutsuka, 2005; Srivastava and Dwivedi, 2006, Jian {\it et al.,} 2009, and references cited therein).
The second question is crucial at present.
The consensus of solar-wind research, however, is unclear as to the origin of the 
mass supply to the supersonic wind. Tian {\it et al.} (2010) have found upflows in the open field-lines of
coronal hole starting in the solar transition region and interpreted this as an
evidence of the 
fast solar wind in the polar coronal holes.
As far as the slow solar wind is concerned, the outflows at the boundaries of active regions
can contribute at larger spatio--temporal scales to the mass supply as an expansion of the loops 
lying over these active regions (Harra {\it et al.,} 2008). It has been shown recently that collimated
jet eruptions can also contribute to the formation of the solar wind (Madjarska, 2013). The magnetic-field topology of 
structures such as jets ({\it e.g.} spray surges) may
not contribute to the solar wind (Uddin {\it et al.,} 2012). The contributions from the confined ejecta
in the solar-wind formation depend mainly on the local magnetic-field topology and 
plasma conditions.

The quest continues on whether the mass supply to the slow solar wind comes from the lower
atmosphere expanding along the curved coronal fields. The question is what are the potential physical
drivers? It has been found that the open field lines at the boundary of active regions
reconnect periodically with closed field lines to guide the plasma motion in the form of
solar wind (Harra {\it et al.,} 2008). Similar examples are reported at the boundary of the
coronal holes at small spatial scales as the source of slow solar wind (Subramanian, Madjarska, and Doyle, 2010).
Therefore, an alternative option may be magnetic reconnection as a potential mechanism in the formation of slow solar wind.
Apart from magnetic reconnection, the wave-heating scenario
 can shed light on the slow solar-wind source regions. Schmidt and Ofman (2011)
have found that the energy stored in the slow magnetoacoustic waves propagating towards the higher atmosphere within expanding loops. This may be a potential candidate for the acceleration 
and formation of the slow solar wind. Wave activity at the bases
of the fast (polar coronal holes) and slow (equatorial corona) solar wind can be important to power the energized plasma at greater heights up to
the corona where it can be triggered supersonically in interplanetary space 
({\it e.g.} Harrison {\it et al.,} 2002; Dwivedi and Srivastava 2006, 2008; De Pontieu {\it et al.,} 2007; McIntosh {\it et al.,} 2011, McIntosh 2012,
and references cited therein). 
Thus, there seems to be compelling evidence for the role of magnetic reconnection and
wave phenomena in the solar wind source region.

In the present article, we report the evidence of the outflowing magnetic arches acting as coronal funnels
at the eastern boundary of an AR 10940 loop system observed on 5 February 2007. These coronal funnels seem to open up 
in the higher atmosphere to transport the outflowing plasma. Their footpoints are rooted
in the boundary of the active region. They are the most likely heated regions that result in the activation of the outflowing plasma. 
We present a 2D MHD simulation of the open and expanding funnel-type
model atmosphere in which a rarefied and hot region is implemented near the footpoint that 
exhibits plasma perturbations similar to the observations. 
In Section~2 we summarize the observational results. We present the numerical model in Section~3.
Discussion and conclusions are given in the last section.

%
\begin{figure*}
\begin{center}
\mbox{
\hspace{-2.0cm}
\includegraphics[width=10.00cm,angle=0]{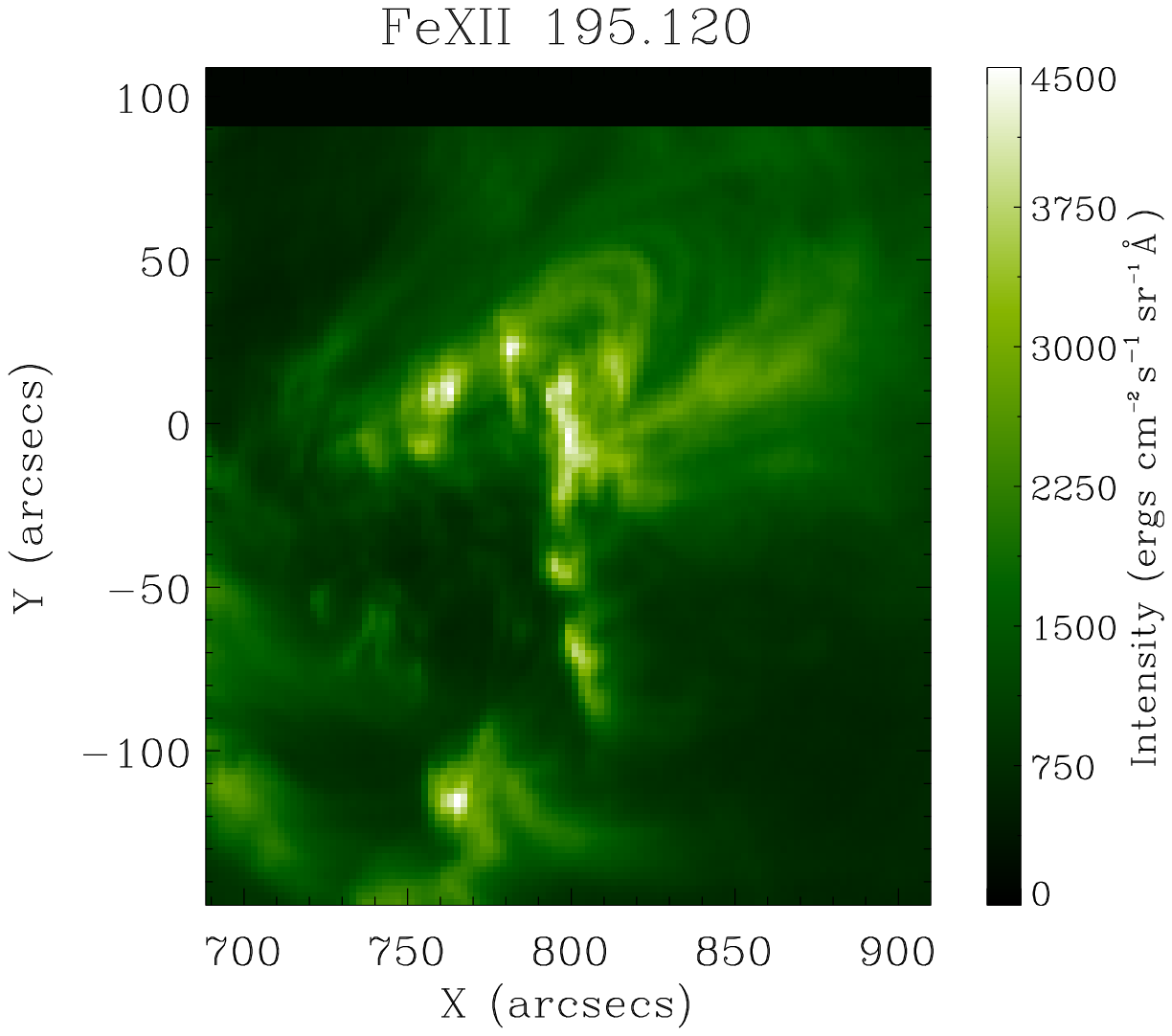}
\hspace{-2.5cm}
\includegraphics[width=10.00cm,angle=0]{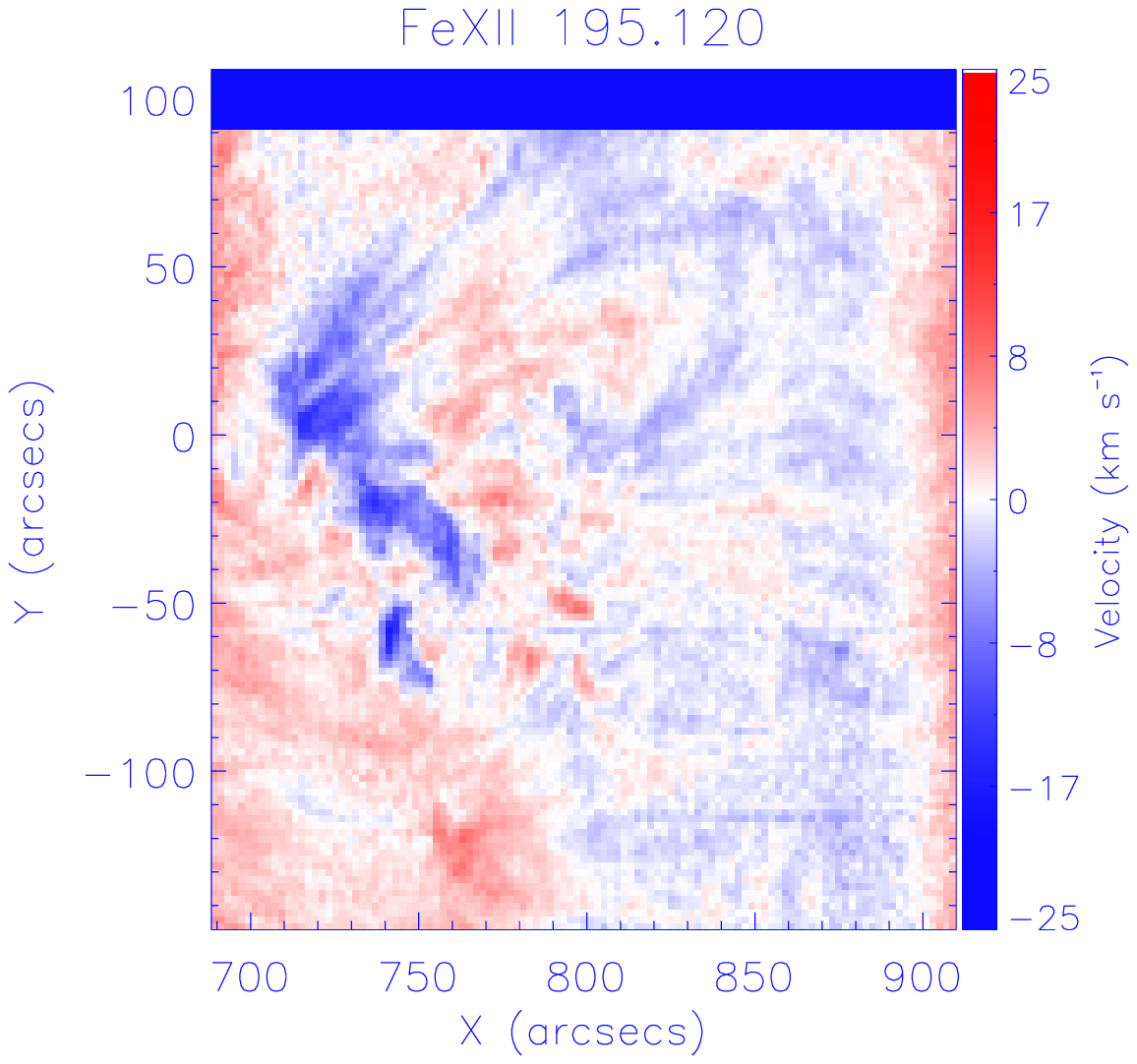}}
\caption{\small Intensity (left) and Doppler velocity (right) maps of Active Region 10940 observed on 5 February 2007. 
The core loops and the open arches at its eastern boundary are clearly evident.
}
\label{fig:obs-int}
\end{center}
\end{figure*}
%
%
\begin{figure*}
\begin{center}
\includegraphics[width=8.00cm,angle=0]{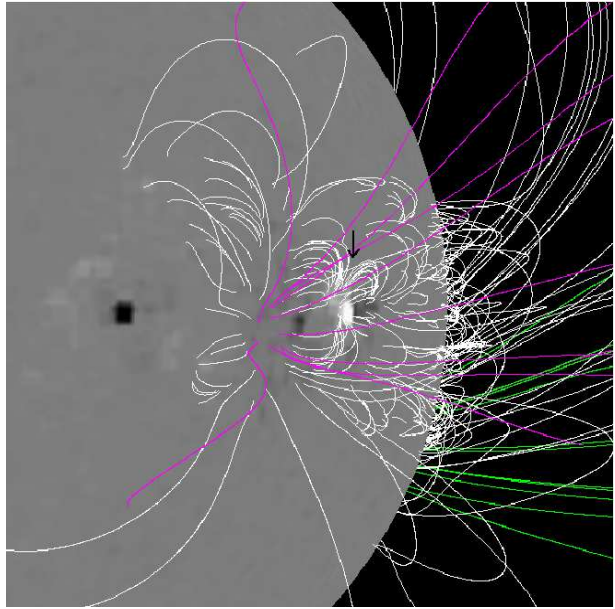}
%
\caption{\small A view of the magnetic-field polarities of AR 10940
and sorrounding regions as observed by {\it Solar} {\it and} {\it Heliospheric} {\it Observatory} {\it (SOHO)}/{\it Michelson} {\it Doppler} {\it Imager} (MDI), and its 
Potential Field Source Surface Extrapolation (PFSS).
The black arrow indicates the core loop system in the AR while the 
south--east part of its boundary shows the large-scale open-field lines
extended into the upper corona. Some of these lines also connect to
the coronal hole lying in the North of AR.
}
\label{fig:obs-int}
\end{center}
\end{figure*}
%
%
\section{Observational Results}
%
\begin{figure*}
\begin{center}
\hspace{-4.0cm}
\includegraphics[width=13.00cm,angle=0]{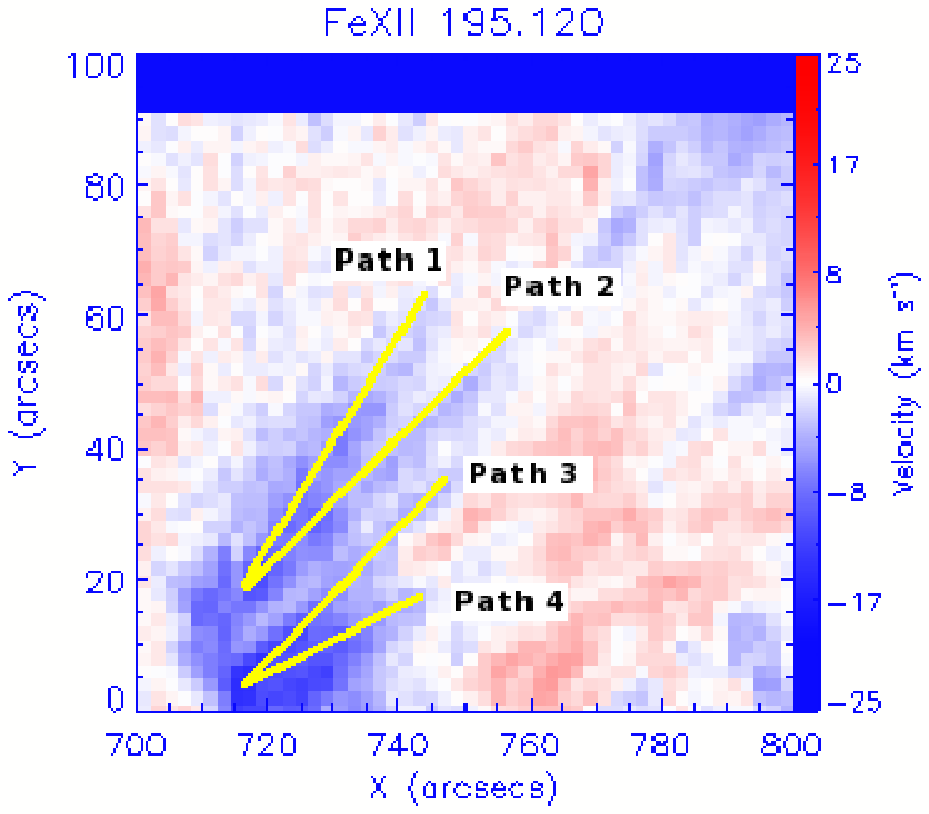}
\mbox{
\hspace{-4.0cm}
\includegraphics[width=8.00cm,angle=0]{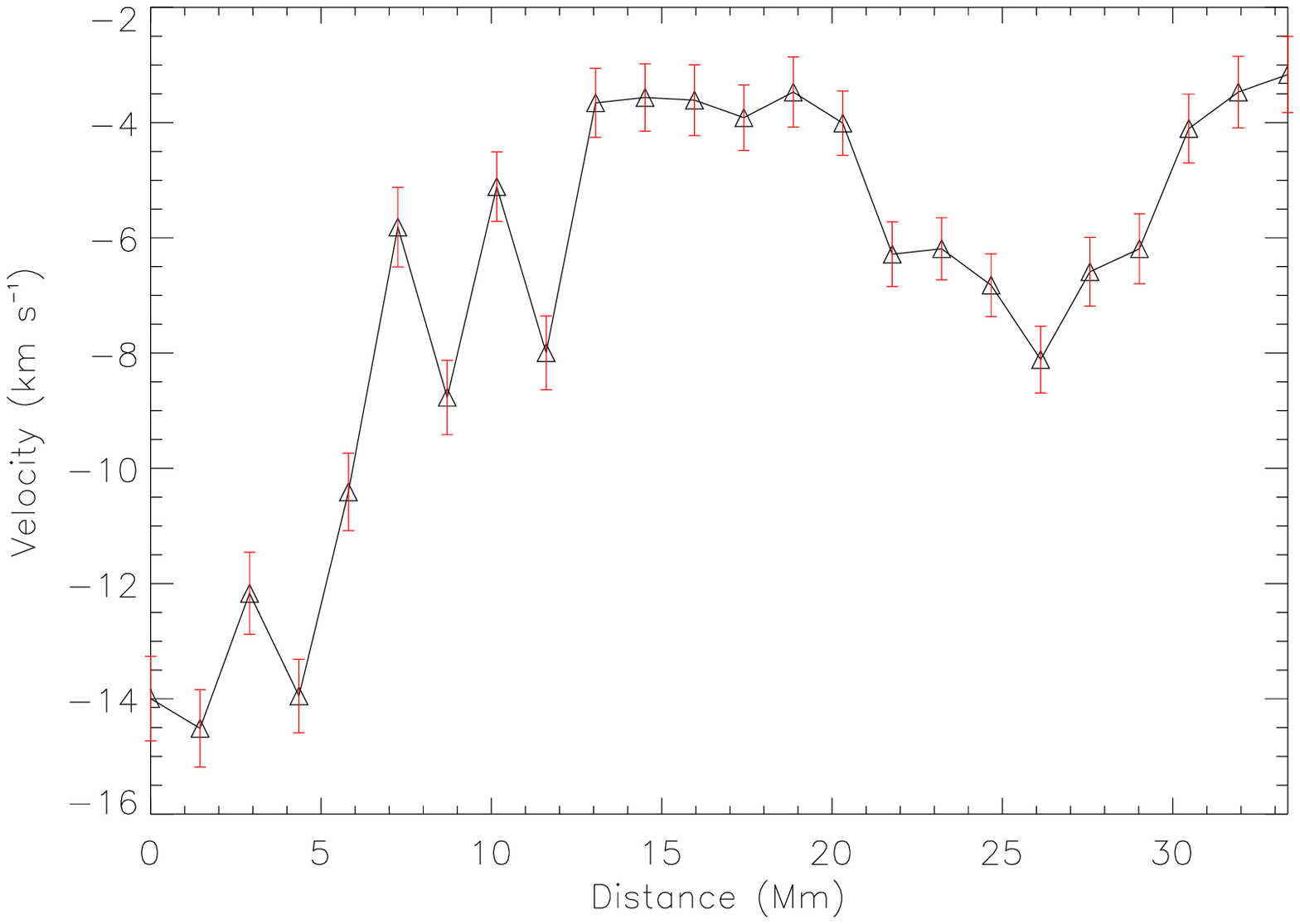}
\hspace{-0.5cm}
\includegraphics[width=8.00cm,angle=0]{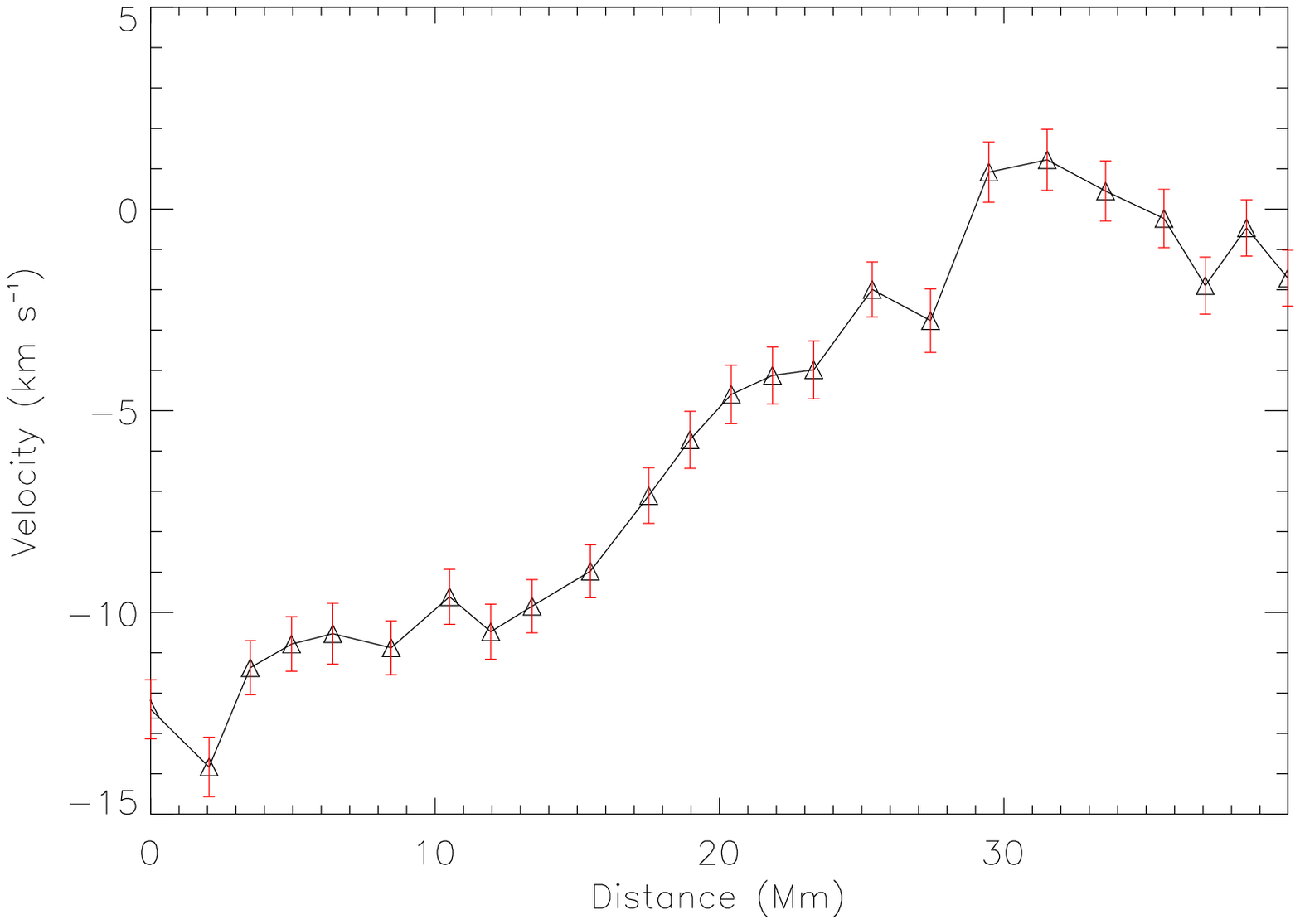}}
\mbox{
\hspace{-4.0cm}
\includegraphics[width=8.00cm,angle=0]{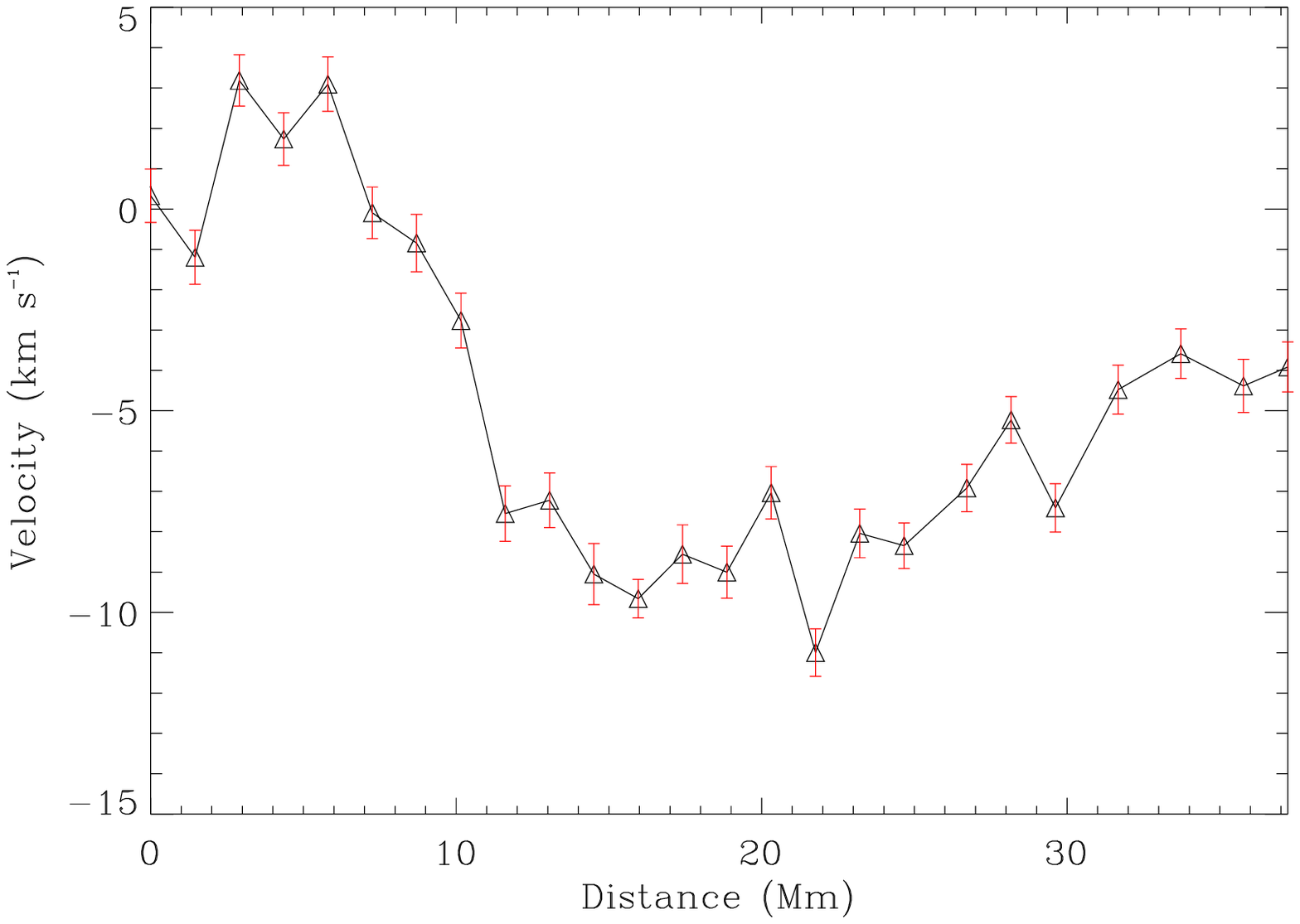}
\hspace{-0.5cm}
\includegraphics[width=8.00cm,angle=0]{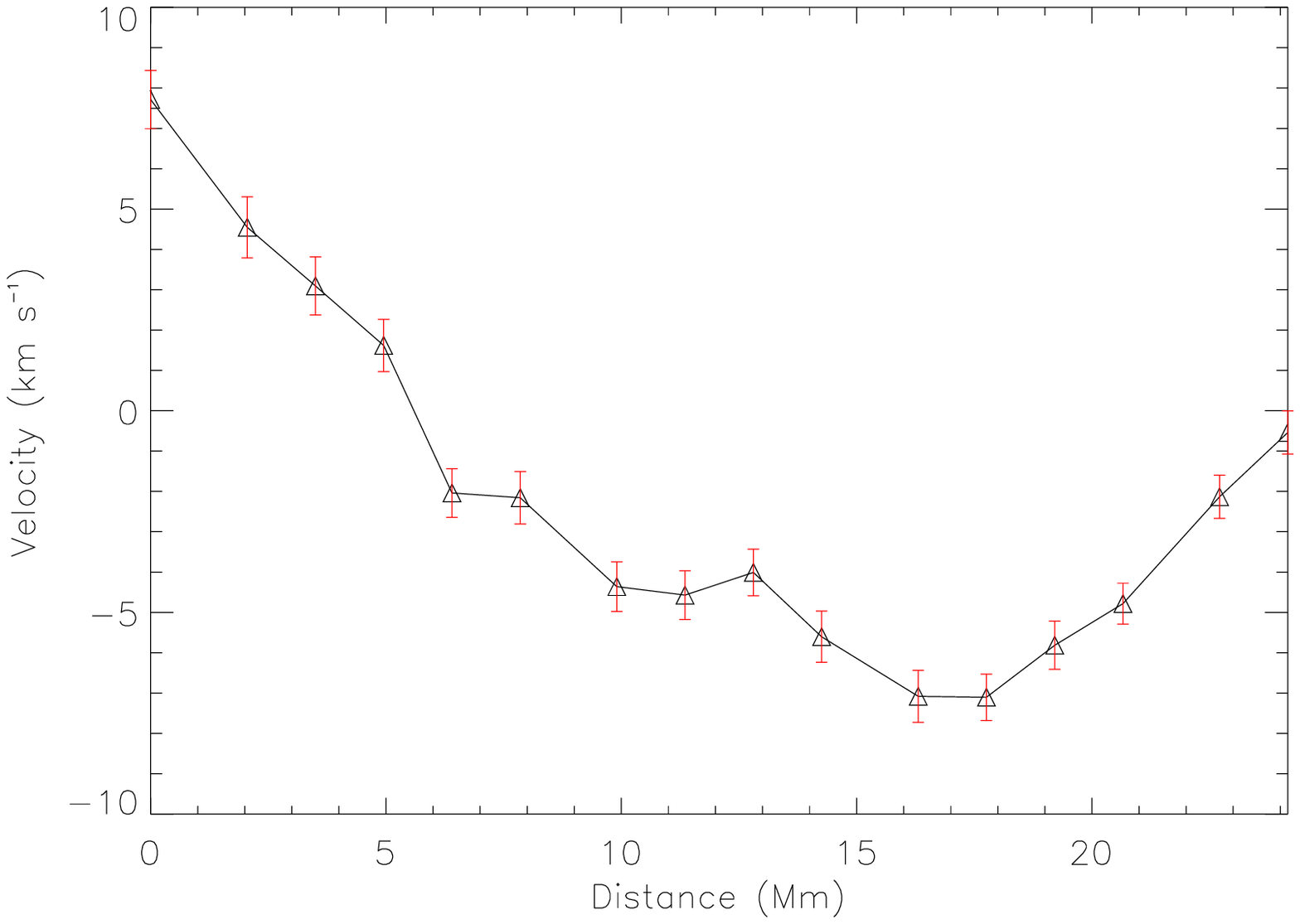}
}
\caption{\small Top: the enlarged velocity map of the eastern 
boundary of AR 10940 showing the open and expanding funnels
in which the plasma is outflowing. Middle and Bottom: the variation of L.O.S.
Doppler velocities along funnels 1 to 4. 
}
\label{fig:obs-vel}
\end{center}
\end{figure*}
The active region AR 10940 was observed in the one-arcsecond-slit scan of {\it EUV Imaging Spectrometer} (EIS: Culhane {\it et al.}, 2006)
onboard the {\it Hinode} spacecraft on 05 February 2007. The EIS is an imaging spectrometer of which 40- and 266-arcsecond slots are used for the image analyses
using the light curves and emissions per pixel. The one- and two-arcsecond slits are utilized for spectral and Doppler analyses
using spectral-line profiles. EIS observes in two modes : i) scan; ii) sit-n-stare.
The EIS observes high-resolution spectra in two wavelength intervals :
170\,--\,211 \AA\ and 246\,--\,292 \AA~using respectively its short-wavelength (SW) and long-wavelength (LW) CCDs. 
The spectral resolution of the EIS is 0.0223 \AA~per pixel. The analysed observations well taken on 05 February 2007 
and the data-set contains spectra of 
various lines formed at chromospheric, transition region (TR), and coronal temperatures.
The scanning observation started at 12:14:12 UT and ended at 13:31:21 UT on 05 February 2007. 
The scanning steps were without any off-set in the region containing a coronal active region and 
its eastern boundary with open and expanding magnetic arches where we are interested in the present investigation. This provides us an opportunity
to understand the plasma activity along the open-field regions at the eastern boundary of the active region in between
the diffused loop systems that reach up to a higher height
in the corona (Figure 1). We refer to these structures as $"$coronal funnels$"$. Such flow regions are heated at their base exhibiting outflows.
To understand the approximate magnetic-field geometry associated with AR 10940 and its sourrounding 
region, we perform a Potential Field Source Surface (PFSS) extrapolation. Fan lines 
over the {\it Solar and Heliospheric Observaory (SOHO)/Michelson Doppler Imager} (MDI) observations are shown in Figure 2. It is clear that the core loops of the active region
are bipolar and connect the central opposite magnetic polarities (white loops shown by the black arrow). Another
set of closed-field lines connect to the east-side weak negative polarity and the central positive polarity. At the eastern 
boundary, there are weak patches of positive and negative polarities from where the open-field lines extend up to higher corona. These open-field lines form the large-scale coronal funnels at the eastern boundary of the active region. The lower parts
of the loops at the eastern boundary can also serve as funnels up to a certain height in the corona. The white lines represent
closed magnetic fields, while magenta and green lines represent the open fields that reach at the source surface, having opposite polarities. The funnels and the south-east boundaries of the AR are associated with the large-scale open field lines (magenta). They may also be the lower parts of the closed-loop system (white lines).

In order to obtain the
velocity structures in such coronal funnels, we select the strongest
EIS line, Fe \textsc{xii} 195.12 \AA\ in our study. We aim for the understanding of the
impulsively generated plasma outflows in such funnels and associated physical
processes. The slit step started the scanning of the polar coronal
hole with ($X_{\mathrm{cen}},Y_{\mathrm{cen}})$$\approx$(799.087 arcsecond, -19.185333 arcsecond). 
The observational windows acquired on the CCDs are 128 pixels high 
along the slit, while 111 pixels wide in the horizontal direction where spectra also disperse 
with the spectral resolution of 0.0222 \AA per pixel.

\begin{figure*}
\begin{center}
\mbox{
\hspace{-4.0cm}
\includegraphics[scale=0.45, angle=0]{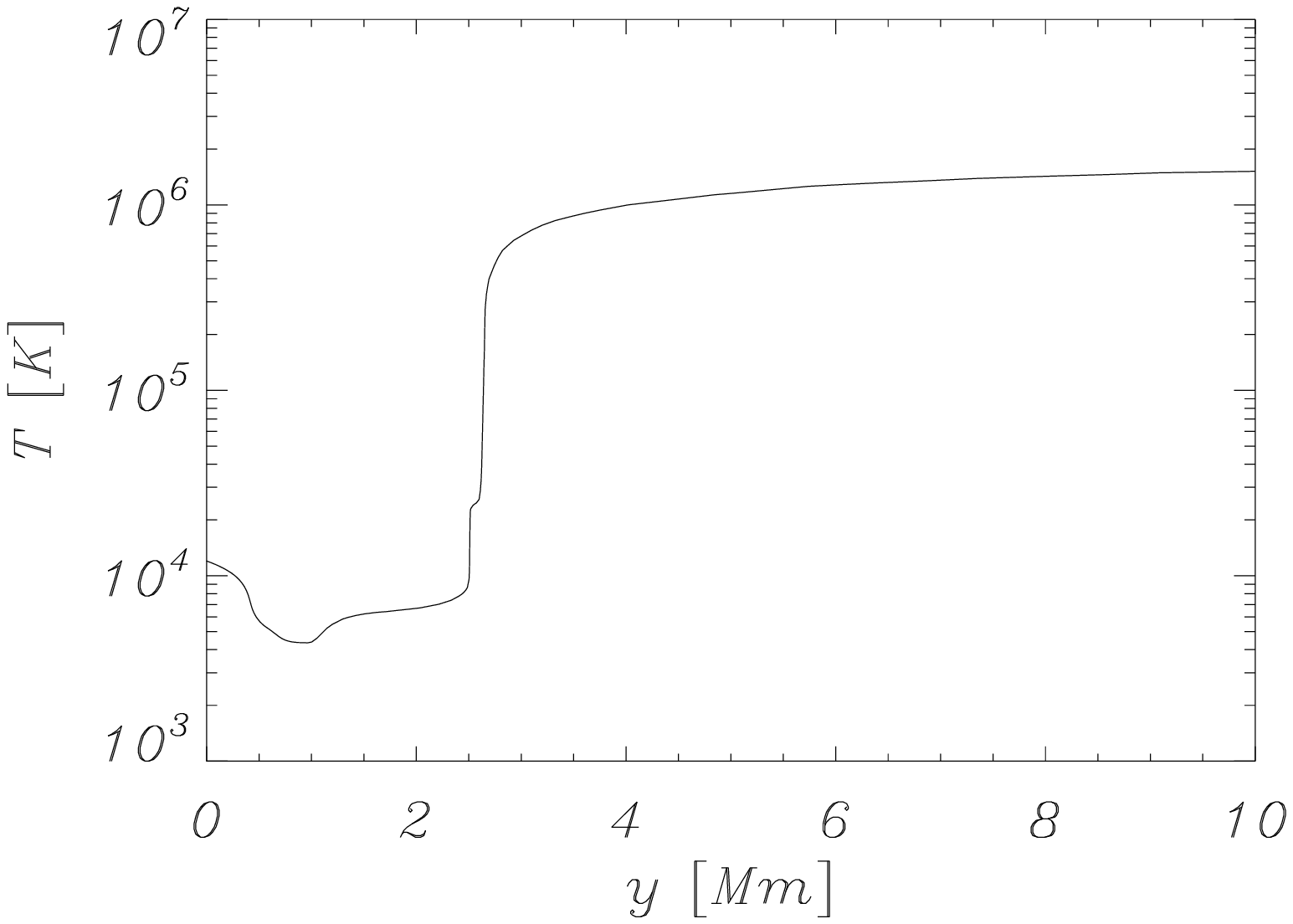}
\hspace{-1.0cm}
\includegraphics[scale=0.45, angle=0]{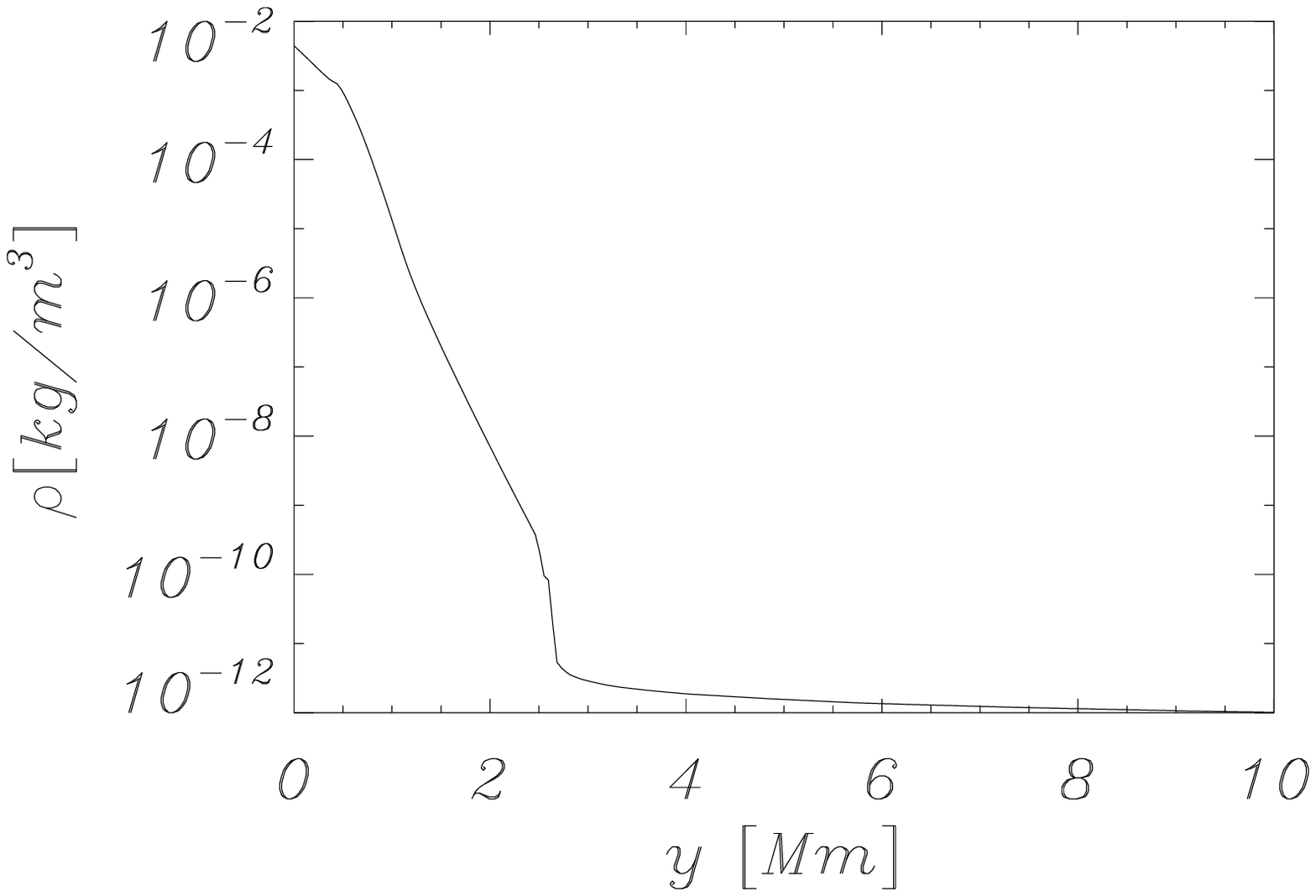}
}
\hspace{-4.0cm}
\includegraphics[scale=0.45, angle=0]{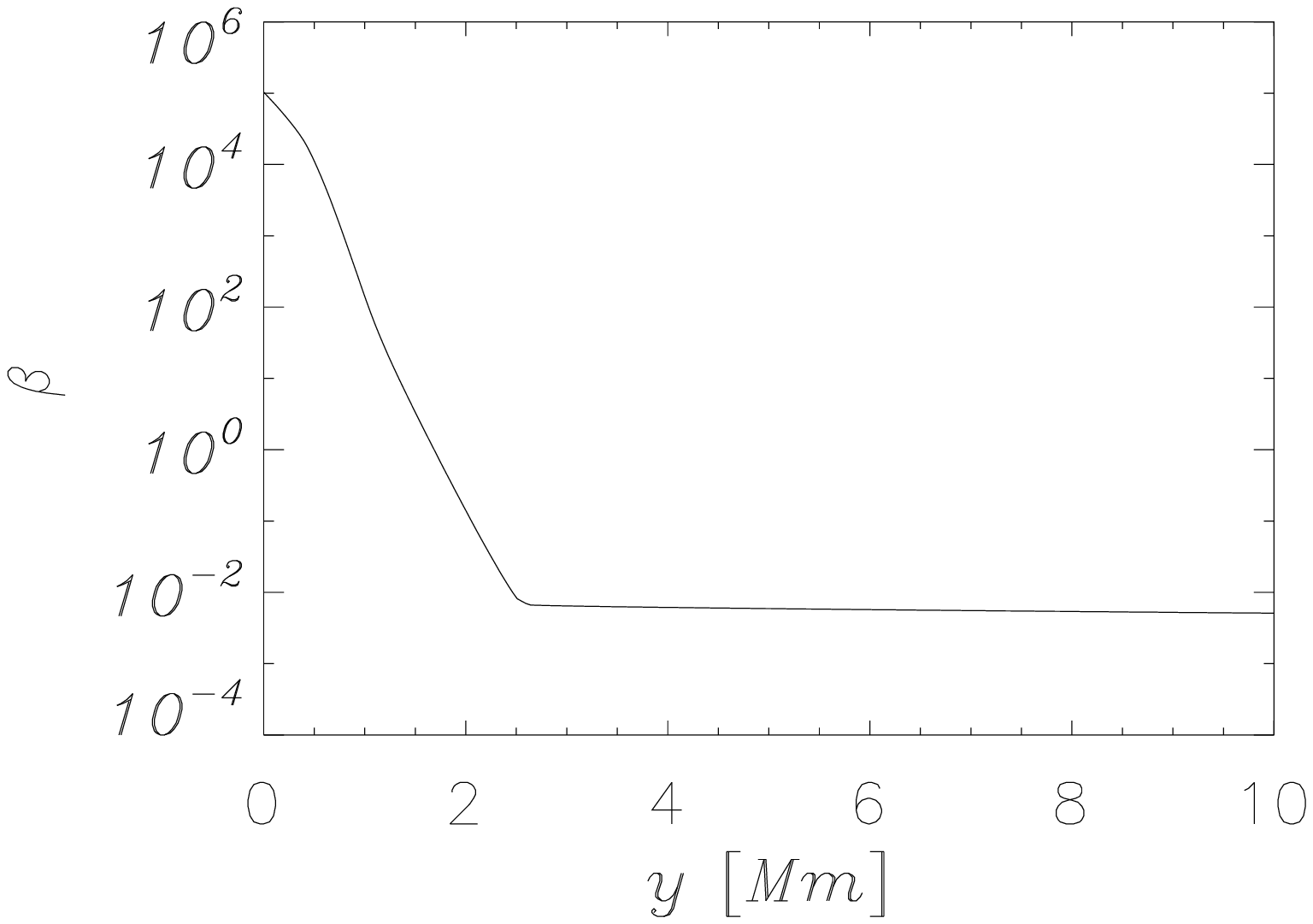}
\caption{\small
Equilibrium profiles of temperature (top-left panel), mass density (top-right panel),
and plasma-$\beta$ (bottom panel) along the vertical line [$y$] for $x=0$. 
}
\label{fig:initial_profile}
\end{center}
\end{figure*}
%
\begin{figure}
\begin{center}
\includegraphics[scale=0.5, angle=0]{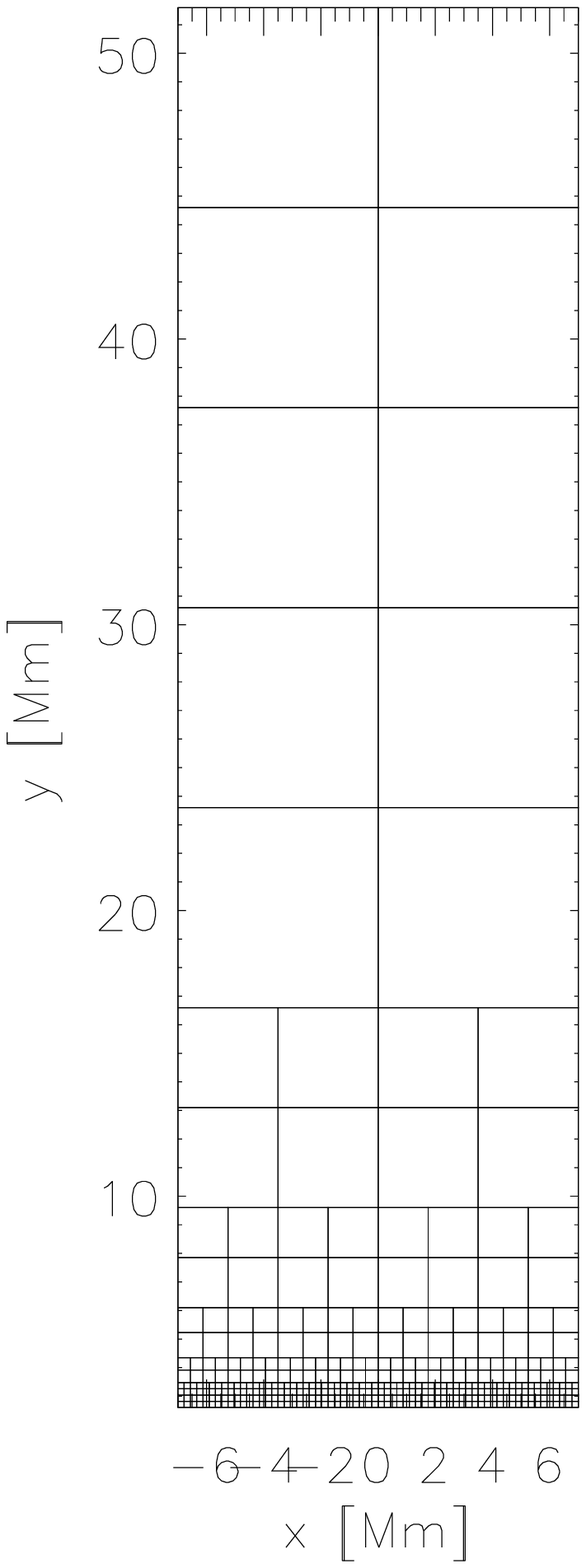}
\vspace{-1.0cm}
\caption{\small Numerical blocks with their boundaries (solid lines). 
}
\label{fig:blocks}
\end{center}
\end{figure}

We apply the standard EIS data-reduction procedures and calibration files/routines
to the data obtained from 
the EUV-telescope, which is the raw (zeroth-level) data. The subroutines are found 
in the {\sf sswidl} software tree under the IDL environment ({\sf www.darts.isas.jaxa.jp/pub/solar/ssw/hinode/eis/}).
These standard subroutines reduce the dark-current 
subtraction, cosmic-ray removal, flat-field correction, hot pixels, warm pixels, and
bad/missing pixels. The data are saved in the level-1 data file, and the 
associated errors are saved in the error file.
We choose the clean and strong line Fe \textsc{xii} 195.12 \AA~to examine the spatial variations
of the intensity, Doppler velocity in the observations scanned over AR 10940 on 5 February 2007.
We co-align the Fe \textsc{xii} 195.12 \AA~map with respect to the long-wavelength CCD observations
of He \textsc{ii} 256.86 \AA~by considering it as a reference image and by estimating
its offset. The orbital and slit-tilt are also corrected in the data using
the standard method described in the EIS-software notes.
We perform a double Gaussian fit for the removal of the weak blending
of Fe \textsc{xii} 195.18 \AA~line as per the procedure described by Young {\it et al.} (2009), 
which is also  outlined in the EIS Software Note 17.
We constrain the weakly blended line 195.18 \AA~to have the same width as the line at 195.12 \AA~and
to fix an offset of +0.06 \AA~relative to it. It is be noted that 
we perform the fitting on 2$\times$2 pixel$^{2}$ binned data to enhance
the signal-to-noise ratio and also to be obtain a reasonable fit. 

Figure~1 displays the intensity (left) and Doppler-velocity (right) maps
of the observed AR and open arches at its eastern boundary.
The intensity map shows that the AR is made up of
the core diffused loop systems lying at the lower heights
in the perpendicular plane to the line-of-sight. In actuality, however, they
may be tilted. These low-lying core loops exhibit high emission and downflows.
At the eastern boundary, we have identified the open and expanding arches
that exhibit coronal funnel behaviour to transport mass and energy
into the higher corona.
These expanding funnels may be the part of more quiescent and large-scale
loop systems opening higher up in the corona.
It is seen in the enlarged Doppler map (see Figure 3) that four such regions are
at least identified as blue-shifted and outflowing regions. As already noted, these could be the legs
of higher and large-scale loops, which form the building blocks
of the coronal funnels at relatively lower heights of the corona. They exhibit the slow and sub-sonic plasma outflows
mostly spreading along their field lines with a maximum speed
of about 10\,--\,15 km~s$^{-1}$.

Figure~3 shows the selection of Path 1\,--4\, (top-panel). The first two paths are drawn over
the large-scale open and magnetic-field arches (lines) lying in a QS\,-\,CH
region in the North East of this AR (not shown here). These two regions serve
as the open and expanding funnel regions. Paths 3 and 4 are drawn to the
expanding and blue-shifted regions that can also serve as coronal funnels. However,
they are actually the lower parts of some core-loop systems. On this account, we have
selected two coronal funnels (1 and 2), while the others (3 and 4) serve as funnels but are probably
associated at larger heights with the
curved-loop magnetic field.
The middle and bottom panels (left to right) respectively show the variation of projected
L.O.S. Doppler velocity along Paths 1\,--4\, with height in these funnels.
For funnels 1 and 2, the footpoints exhibit stronger outflows with a
maximum speed of 15 km~s$^{-1}$. At heights beyond 10 Mm, 
the outflow speed weakens in these funnels. This signifies the start of the 
outflows due to heating near the footpoints of these funnels; the outflows weaken 
with the height as we move away from the heating source.
The other two funnels (3 and 4) are likely the legs of the diffused core loops,
exhibiting the outflows at certain lower heights with a maximum
speed of 10 km~s$^{-1}$. This indicates different locations of the 
heating. It is also noted that the outflows
diminish at lower heights compared to the corressponding ones in Funnels 1 and 2 in the form of open arches. This
is because of their association with the curved loops at greater heights where
plasma is trapped and flows downward in order to maintain a new equilibrium.
It is also noted that the flow structure is more gentle in the open Funnels 1 and 2. However, 
it is greater at the footpoint and decreases with height. In the Funnels 3 and 4, which are 
the lower parts of curved loops, the generation of the flow is rather impulsive. The outflow starts 
at a certain height above the loop's footpoint. It increases up to a certain distance and decreases thereafter.
This shows that impulsive heating is at work near the loop footpoint, which causes enhanced 
upflows up to a certain distance. The downflowing plasma from the upper part of the loop  
may counteract with the upflows. The red-shifted apex of the core loop system is clearly evident 
in Figures 1 and 3. Funnels 3 and 4 are the lower parts of this loop system.

In the next section, we outline the details of the 2D numerical
simulation of such observed  expanding coronal funnels and their plasma dynamics. We solve a set of ideal MHD equations 
in the appropriate VAL-III C initial temperature  
conditions and model atmosphere using the FLASH code.

\section{Numerical Model of the Hot Plasma Outflows in Coronal Funnels}
\subsection{Model Equations}
Our model system acquires a
gravitationally stratified solar atmosphere 
which can be described by
the ideal two-dimensional (2D) 
MHD equations:
\begin{equation}
\label{eq:MHD_rho}
\hspace{-3.3cm}
{{\partial \varrho}\over {\partial t}}+\nabla \cdot (\varrho\textbf{\textsl{V}})=0\, ,
\end{equation}
\begin{equation}
\label{eq:MHD_V}
\varrho{{\partial \textbf{\textsl{V}}}\over {\partial t}}+ \varrho\left (\textbf{\textsl{V}}\cdot \nabla\right )\textbf{\textsl{V}} =
-\nabla p+ \frac{1}{\mu}(\nabla\times\textbf{\textsl{B}})\times{\textbf{\textsl{B}}} +\varrho{\textbf{\textsl{g}}}\, ,
\end{equation}
\begin{equation}
\label{eq:MHD_p}
\hspace{-1.3cm}
{\partial p\over \partial t} + \nabla\cdot (p\textbf{\textsl{V}}) = (1-\gamma)p \nabla \cdot\textbf{\textsl{V}}\, ,
\end{equation}
\begin{equation}
\label{eq:MHD_B}
\hspace{1.8cm}
{{\partial\textbf{\textsl{B}}}\over {\partial t}}= \nabla \times (\textbf{\textsl{V}}\times\textbf{\textsl{B}})\, , 
\nabla\cdot\textbf{\textsl{B}} = 0\, .
\end{equation}
Here ${\varrho}$, $\textbf{\textsl{V}}$, $\textbf{\textsl{B}}$, $p = \frac{k_{\rm B}}{m} \varrho T$, $T$, $\gamma=5/3$,
$\textbf{\textsl{g}}=(0,-g)$ with its value $g=274$ m s$^{-2}$, $m$, and $k_{\rm B}$ are  
the mass density, flow velocity, magnetic field, gas pressure, temperature,
adiabatic index, gravitational acceleration, mean particle mass, and Boltzmann's constant respectively.
It should be noted that we do not consider radiative cooling and 
thermal conduction in our present model for the sake of simplicity. We simulate only the dynamics of the
plasma outflows to compare them with that of the observed coronal funnels.
%
%
\begin{figure*}
\begin{center}
\vspace{-0.25cm}
\mbox{
\includegraphics[scale=0.75, angle=0]{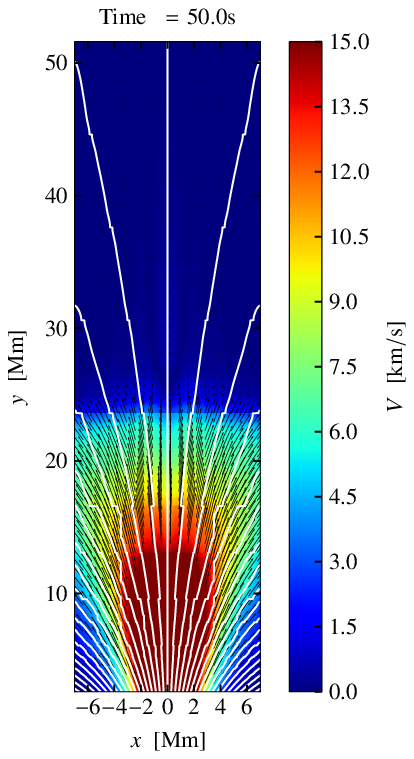}
\includegraphics[scale=0.75, angle=0]{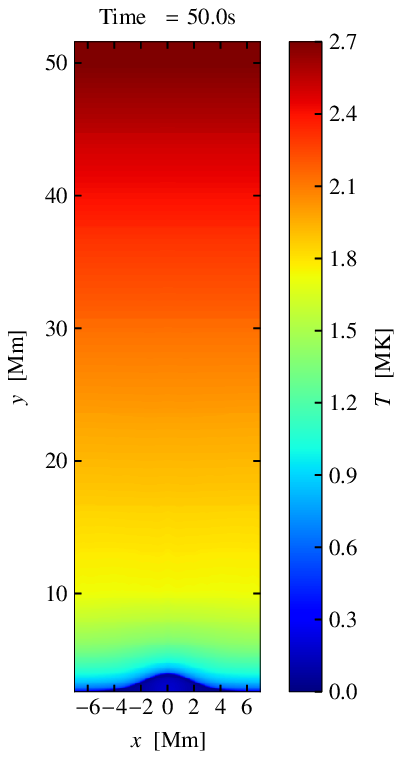}
\includegraphics[scale=0.75, angle=0]{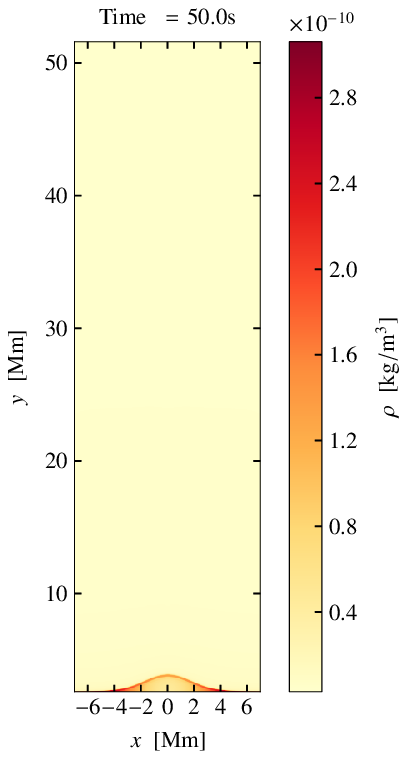}
}
\vspace{-0.50cm}
\mbox{
\includegraphics[scale=0.75, angle=0]{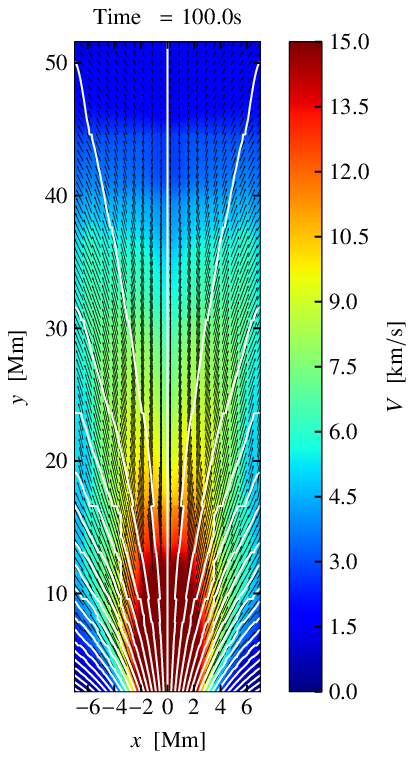}
\includegraphics[scale=0.75, angle=0]{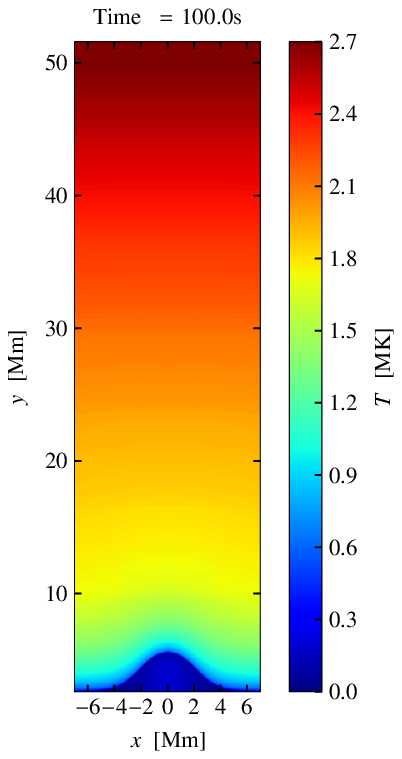}
\includegraphics[scale=0.75, angle=0]{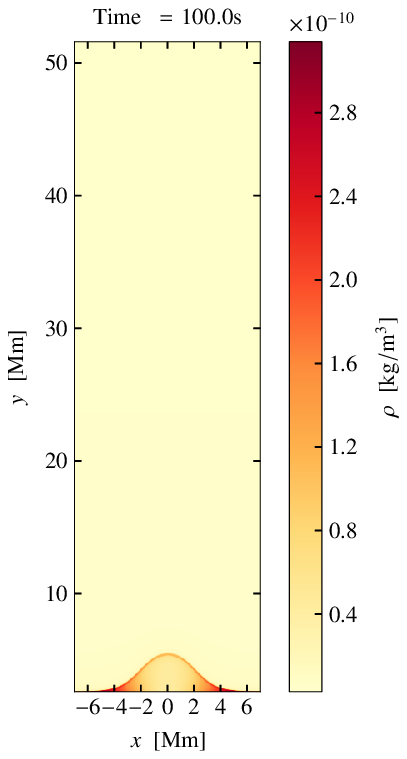}
}
\vspace{-0.50cm}
\mbox{
\includegraphics[scale=0.75, angle=0]{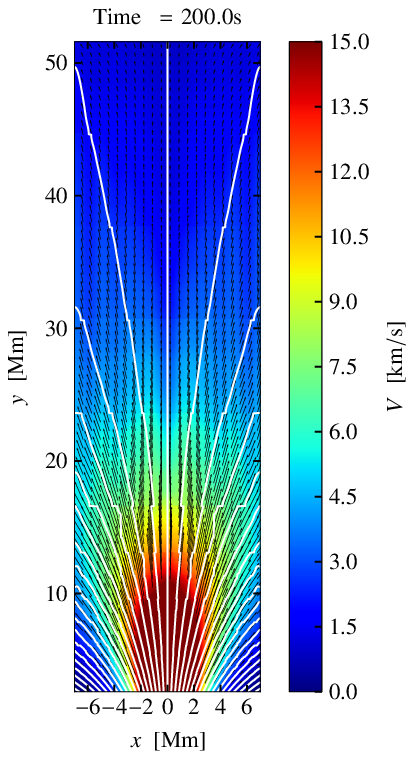}
\includegraphics[scale=0.75, angle=0]{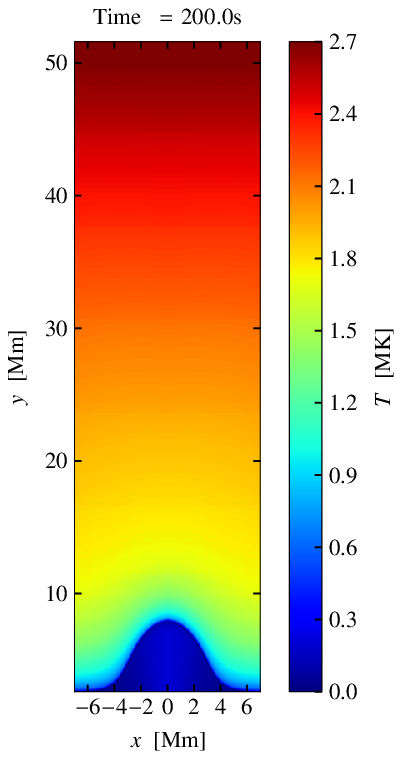}
\includegraphics[scale=0.75, angle=0]{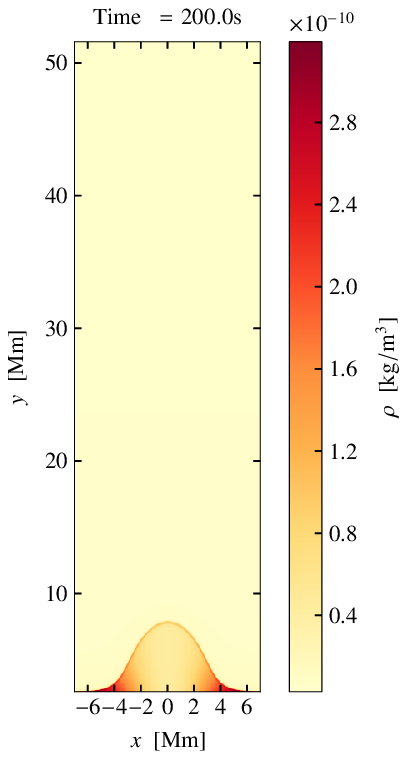}
}
\vspace{-0.50cm}
\mbox{
\includegraphics[scale=0.75, angle=0]{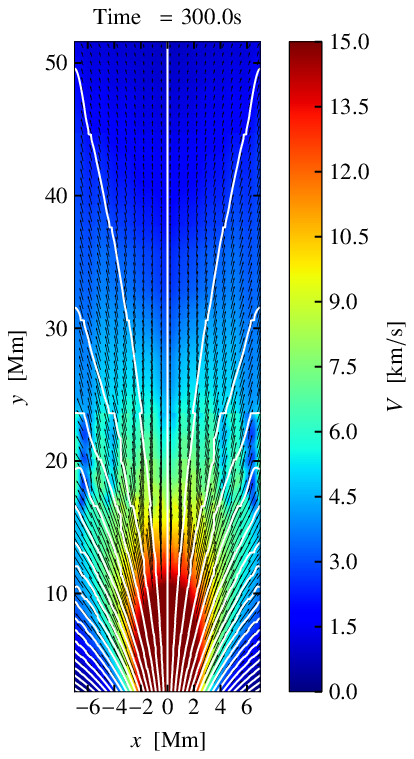}
\includegraphics[scale=0.75, angle=0]{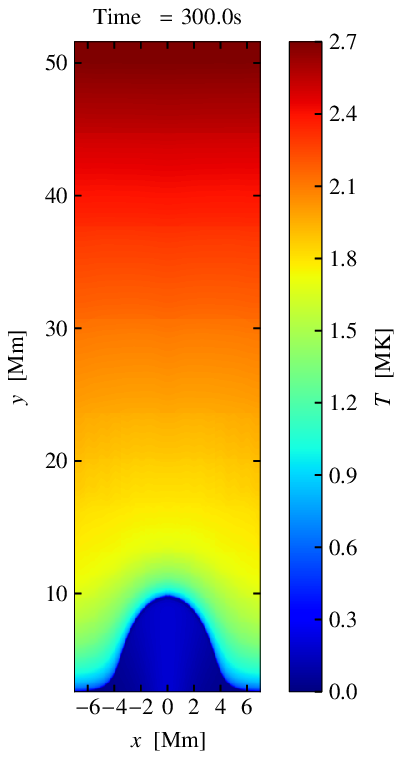}
\includegraphics[scale=0.75, angle=0]{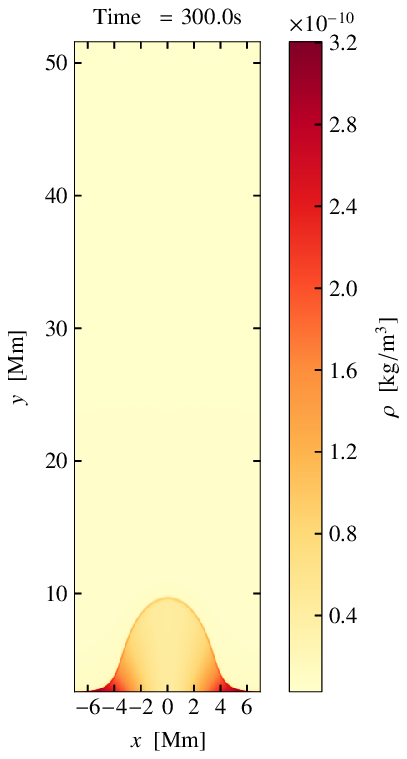}
}
\vspace{0.50cm}

\caption{\small Numerical results (top to bottom rows): the velocity vectors superposed over the total velocity (left); the temperature (middle) and density (right) maps, for {\it t}\,=\,50, 100, 200, 300 seconds. White lines represent the diverging magnetic-field lines. The complete movies are given as an electronic
supplementary material for these numerical results.
}
\label{fig:num-V}
\end{center}
\end{figure*}
%
%
%
%
%
\subsection {Initial Conditions}
%
%
%
We assume that the solar atmosphere is at rest $\left[\textbf{\textsl{V}}_{\rm e}={\bf 0}\right]$ in equilibrium with a current-free 
magnetic field,
%
%
%
$
\left[\nabla \times {\textbf{\textsl{B}}}_{\rm e}={\bf 0}\right].
$
%
As a result, the magnetic field is force-free
\begin{equation}\label{eq:B_e}
(\nabla\times{\textbf{\textsl{B}}}_{\rm e})\times{\textbf{\textsl{B}}}_{\rm e} = {\bf 0}\ . 
\end{equation}
The divergence-free constraint is satisfied automatically if 
the magnetic field is specified by the magnetic-flux function, $\left[A(x,y)\right]$,
as
\begin{equation}\label{eq:B_e1}
{\textbf{\textsl{B}}}_{\rm e}=\nabla \times (A\hat {\bf z})\, .
\end{equation}
Here the subscript $'e'$ corresponds to equilibrium quantities.
We set a curved magnetic field by choosing
\begin{equation}
A(x,y) = S\frac{x-a}{(x-a)^2+(y-b)^2},
\end{equation}
where
$S$ 
is
the strength of magnetic pole and $(a,b)=(0,-10)$ Mm is its position.
%
%
For such a choice of $(a,b)$, the magnetic-field vectors are weakly curved
to represent the expanding coronal funnels. 

As a result of Equation~(\ref{eq:B_e}) 
the pressure gradient is balanced by the gravity force,
\begin{equation}
\label{eq:p}
-\nabla p_{\rm e} + \varrho_{\rm e} {\bf g} = {\bf 0}\, .
\end{equation}
With the ideal-gas law and the $y$-component of 
Equation~(\ref{eq:p}), we 
arrive at 
\begin{equation}
\label{eq:pres}
p_{\rm e}(y)=p_{\rm 0}~{\rm exp}\left[ -\int_{y_{\rm r}}^{y}\frac{\bf \mathrm{dy^{'}}}{\Lambda (y^{'})} \right]\, ,\hspace{3mm}
\label{eq:eq_rho}
\varrho_{\rm e} (y)=\frac{p_{\rm e}(y)}{g \Lambda(y)}\, ,
\end{equation}
where
\begin{equation}
\Lambda(y) = \frac{k_{\rm B} T_{\rm e}(y)}{mg}
\end{equation}
is the pressure scale height, and $p_{\rm 0}$ denotes the gas 
pressure at the reference level that we choose in the solar corona at $y_{\rm r}=10$ Mm.

We take
an equilibrium temperature profile $\left[T_{\rm e}(y)\right]$ (see top-left panel in Figure~4) for the solar atmosphere
derived from the VAL-C atmospheric model of Vernazza, Avrett, Loeser (1981) that is smoothly extended to the solar corona. 

The transition region 
is located at $y\approx 2.7$ Mm. 
There is an extended solar corona above the transition region, 
and with the temperature minimum level located at $y\approx0.9$ Mm below the solar chromosphere.
Having specified $T_{\rm e}(y)$ (see Figure~4, left-top panel) with Equation~(\ref{eq:pres}), 
we obtain the corresponding gas-pressure and mass-density profiles.

%
%
%
%
%
\begin{figure*}
\begin{center}
\mbox{
\includegraphics[scale=1.25, angle=0]{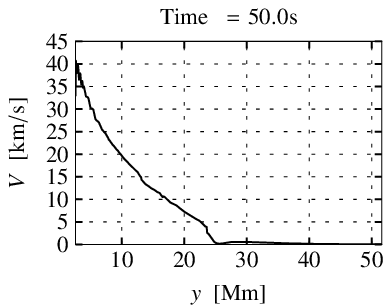}
\includegraphics[scale=1.25, angle=0]{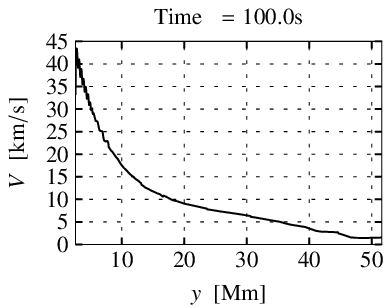}
}

\mbox{
\includegraphics[scale=1.25, angle=0]{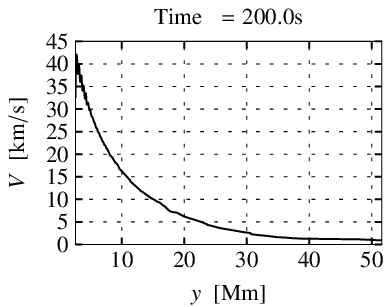}
\includegraphics[scale=1.25, angle=0]{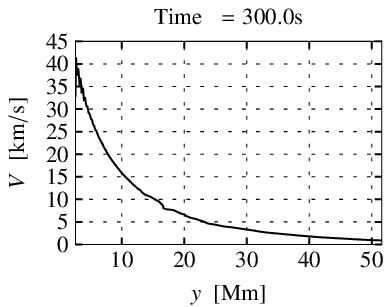}
}



\caption{\small
Velocity profiles plotted w.r.t the height along the model funnel at
50, 100, 200, 300 seconds.}
\label{fig:num-V}
\end{center}
\end{figure*}
%
%
%

\subsection{Numerical Scheme and Computational Grid}
Equations (\ref{eq:MHD_rho})\,--\,(\ref{eq:MHD_B}) are solved numerically using the FLASH code
(Lee and Deane, 2009; Lee, 2013). This code uses a second-order unsplit Godunov solver (Godunov, 1959)
with various slope 
limiters and Riemann solvers, as well as adaptive mesh refinement (AMR) (Murawski and Lee, 2011, 2012).
We set the simulation box 
of $(-7,7)\, {\rm Mm} \times (2.6,51.6)\, {\rm Mm}$ along the $x$- and $y$-directions 
(Figure~5). The lower boundary, where we apply the heating pulse, is set at 
$x_{\rm 0}=0$ Mm, $y_{\rm 0}=2.6$ Mm. We set and hold fixed all of the plasma quantities at all boundaries of the simulation region to their equilibrium values 
which are given by Equations~(\ref{eq:B_e1}) and (\ref{eq:eq_rho}).
As the magnetic field is curved and plasma is stratified gravitationally, open boundaries would not be a perfect choice with regard to the fixed boundary conditions. We have verified this experimentally. As the FLASH code uses a third-order accurate Godunov-type method, a characteristic method is already built into this procedure. However, the Riemann problem at the boundaries  corresponds to open boundaries, resulting in numerically induced reflections from these boundaries. On the other hand, the fixed boundaries lead to negligibly small numerical reflections. Therefore, we adopt this approach in the numerical simulations.
In addition, we modify the equilibrium mass density and gas pressure at the bottom boundary as
\begin{equation}
\varrho(x,y,t) = \varrho_{\rm e}(y)
\left[
1 + A_{\rm \varrho} f(x,y) g(t)
\right]
\, ,
\label{eq:perturb}
\end{equation}
\begin{equation}
p(x,y,t) = p_{\rm e}(y)
\left[
1 + A_{\rm p} f(x,y) g(t)
\right]
\, ,
\label{eq:perturb}
\end{equation}
where
\begin{equation}
f(x,y) = 
\exp\left[ 
-\frac{(x-x_{\rm 0})^2} {w_{\rm x}^2}
-\frac{(y-y_{\rm 0})^2} {w_{\rm y}^2} 
\right]
\end{equation}
and
\begin{equation}
g(t) = -
\left[ 
\exp\left(
-\frac{t}{\tau}
\right) - 1
\right]
\, .
\end{equation}
Here $A_{\rm \varrho}$ and $A_{\rm p}$ are the amplitudes of the perturbations, $(x_{\rm 0},y_{\rm 0})$ 
are their initial position and
$w_{\rm x}$, $w_{\rm y}$ denote its widths along the $x$- and $y$-directions, respectively. 
The symbol $\tau$ denotes the growth time of these perturbations. 
We set and hold fixed $A_{\rm \varrho}=7$, 
$A_{\rm p}=2$, 
$x_{\rm 0}=0$ Mm, $y_{\rm 0}=2.6$ Mm, 
$w_{\rm x}=3$ Mm, $w_{\rm y}=0.5$ Mm, and $\tau=10$~seconds.
These fixed boundary conditions perform much better than transparent boundaries, 
leading to only negligibly small numerical 
reflections of the wave signals from these boundaries. 

In our modeling, we use an AMR grid with a minimum (maximum) level of 
refinement set to $3$ ($8$) (see Figure~5). The refinement strategy is based on 
controlling numerical errors in 
mass density, which results in an excellent resolution of steep spatial profiles and
greatly reduces numerical diffusion at these locations. 
 
A standard procedure to check the magnitude of a numerically induced flow is to run the code for the equilibrium alone, 
without implementing any perturbation. 
We have verified that the numerically induced flow is of the order of $2$ km s$^{-1}$ in the solar corona, 
and the transition region is not affected by the spatial resolution which is about $20$ km around 
the transition region. 
This resolution is much smaller than the width of the transition region ($\approx 200$ km) 
and the pressure scale height near the bottom, which is about $0.5-1$ Mm. 
We have also performed grid convergence studies by increasing the spatial resolution by a factor of two at the transition 
region. As the numerical results have been found essentially similar results for the finer and coarser grid, we have limited 
our analysis to the latter.

\subsection{Results of the Numerical Simulation and Comparison with Observations}

The results of the numerical simulations and their comparison with the observed plasma outflows are summarized as follows:

Figure~6 (first column) displays the velocity vectors plotted over the total velocity maps for $t=50$, 
$100$, $150$, $200$, $250$, $300$, $350$, $400$, $450$ seconds. 
The diverging magnetic-field lines of the model funnel is also over-plotted
in these snapshots. It is clear from the $t=50$ seconds snapshot that 
the implemented heating at the footpoint, just 
below the transition region, results in the alteration of the ambient plasma pressure, and the plasma starts
flowing upward with a typical average speed of 40 km s$^{-1}$ near the footpoint at this time. The outflow velocity
is maximum near the heated region at the footpoint of the model funnel. This is higher than the observed line-of-sight outflow velocities (10\,--\,16 km s$^{-1}$) in various coronal funnels above their footpoints (see Figure~3).
The simulated outflow velocities greatly depend on the initial conditions of 
the model funnel and magnitude of the heating pulse.
The observed trend of the flows in Funnel 1 and 2 (see Figure 3) and the same derived from the model match with each other (see Figures 6 and 7). At higher altitudes above the heating location in these funnels, the outflow weakens
which qualitatively match with the results obtained from our model (see Figures 6 and 7). Our model  better fits the plasma flow conditions in the open coronal funnels ({\it e.g.} funnels 1 and 2) where the gentle flows start and expand due to their footpoint heating.
The flow becomes steady at each height after 300 seconds in the model funnel when the 
heated plasma reach the height of 10 Mm. This also indicates that
the heating pulse is at work for a certain duration, and after some time the generated flows reach
a new equilibrium. The physical scenario is in agreement with the open coronal Funnels 1 and 2
(see Figures 3 and 7). Comparison of these two scenarios supports the outflow of the plasma due to heating. The heating causes
thermal flows guided in the magnetic-field lines of the open coronal funnels while it subsides away from the source. 
The physical behaviour of the velocity field and its spatial distribution as observed by Hinode/EIS along each funnel
are consistent with the velocity field at a particular temporal span of the numerical simulation when 
the outflowing plasma rise to a maximum height of $\approx$10 Mm. 

For the funnels
that are the lower parts of the curved loop system (Funnels 3 and 4; Figure 3), the energy release might occur where blue-shift is enhanced at a certain height above the footpoint impulsively. Therefore, the locations in those observed funnels are identical to the energy release site of the modeled funnel where outflows start as a result of heating.  On the contrary, it seems that impulsive heating in Funnels 3 and 4 causes the increment in the outflows up to a certain height and thereafter is balanced by 
the downflowing and trapped plasma from the loop apex.

Figures~6, second and third columns, respectively display the temperature and density maps for $t=50$, 
100, 200, 300 seconds. It is clear that during the heating, the
plasma maintained at inner coronal/TR temperatures (sub-MK and 1.0 MK; the hot plasma envelops the cool one) and with somewhat higher density, starts flowing from the footpoint
of the model funnel towards greater heights. The plasma is denser
near the footpoint. It flows streamline along the open funnels towards higher
heights. Thermal perturbations created the slow and subsonic flows
of the plasma; therefore, it reaches up to only the lower heights. 
This is a similar situation that of the transition region within the funnel being pushed upward
due to the evolution of the thermal perturbation underneath.  
Sub-mega-Kelvin and denser plasma from lower solar atmosphere move up and is
enveloped by the hot coronal plasma.
The denser plasma maintained at 
the TR and inner coronal temperatures is visible up to a height of 10 Mm in the 
model funnel. This is consistent with the 
observations. Only the lower parts of the coronal arches are associated with enhanced fluxes and thus
with higher densities, while their higher parts are less intense and denser regions (see Figure 1, left panel). 

\section{Discussion and Conclusions}
We have presented observations of outflowing coronal arches (coronal Funnels 1 and 2) lying at the
eastern boundary of the AR 10940 observed on 5 February 2007, 12:15\,--\,13:31 UT.
The scanning observations show that these arches open up in the nearby
quiet-Sun corona and exhibit plasma outflows maintained at coronal/TR temperature of around 1.0 MK.
They serve as the expanding coronal funnels from which the plasma moves to the
higher corona and serves as a source to the slow solar wind (Harra {\it et al.,} 2008).
The plasma outflows may be generated in such open field regions
because of the low atmospheric reconnections between the open- and closed-
field lines (Subramanian {\it et al.,} 2010). Episodic heating mechanisms 
are now well observed and interpreted as the drivers of 
outflowing plasma in the curved coronal loops (Klimchuk 2006; Del Zanna 2008; Brooks and Warren 2009).
The steady heating may generate the hot-plasma upflows in the corona (Tripathi {\it et al.,} 2012).

There has so far been little effort to model the plasma outflows in the corona and that too in an entirely different context
of the large-scale evolution of coronal magnetic field. 
Murray {\it et al.} (2010) have modeled in 3D the origin and driver of coronal outflows, 
and found that outflows are the result of the expansion of an active region during its development.
Harra {\it et al.} (2012) have modeled the AR coronal outflows as a consequence of compression 
during the creation and annihilation of the magnetic-field lines.

However, our objective of the present investigation is to implement and test our 2D numerical
simulation model of the localized expanding coronal funnels with 
{\it Hinode}/EIS observations of the outflowing open-field arches ({\it i.e.} funnels) at the 
boundary of AR 10940. We solve the set of ideal MHD equations 
in the appropriate VAL-III C initial temperature  
conditions and model atmosphere using the FLASH code.
The key engredient of our model is the implementation of realistic ambient solar atmosphere, {\it e.g.} 
realistic temperature, presence of the TR, implementation of expanding coronal fields, stratified
atmosphere in the initial equilibrium, which affect significantly such plasma dynamics.
We have implemented a rarefied and hotter region at the footpoint of the model funnel that 
triggers the evolution of the slow and sub-sonic plasma perturbations propagating
outward in the form of plasma flows similar to the observed
dynamics. 
We implemented the localized heating
below the transition region  at the footpoint of the funnel.
The plasma is considered rarefied in the horizontal direction, mimicking
the structured, open, and expanding magnetic funnels. The heated plasma
evolves and exhibits the plasma perturbations as has been observed as 
outlined in the {\it Hinode}/EIS observations. The outflows start at the base 
of the funnel that further weakens with the height, which is suggestive of the
plasma dynamics due to heating near the footpoint. A similar physical
scenario is observed in the selected outflowing magnetic arches mimicking
the coronal funnels 1 and 2 at the eastern boundary of AR 10940.
 
We conclude that the implemented episodic heating 
can excite plasma outflows in the solar active regions. These
slow, subsonic plasma outflows may not be launched at higher altitudes in the corona.
We have examined the presence of hot and denser plasma up to 10 Mm height 
as triggered by thermal perturbations in our model. Observations also 
show that the plasma only rises significantly in the lower parts of these
funnels up to inner coronal heights with significant intensity and velocity distributions. However, even if such flows reach up to
the inner coronal heights of 10 Mm in these funnels, they may also contribute
to the mass supply to the slow solar wind. Therefore, our model and observations invoke
the dynamics of the plasma in the localized coronal funnels, which
may be important candidates to transport mass and energy into the inner corona.
However, more observational studies should be performed
out new spectroscopic data ({\it e.g.} {\it Interface} {\it Region} {\it Imaging} {\it Spectrograph} (IRIS)) to compare with our proposed 2D model, specifically 
under the physical conditions of different types of localized fluxtubes in the solar atmosphere,
which can serve as plasma outflowing regions due to episodic heating.

\begin{acks} 
We thank the referees for their valuable suggestions 
which improved the manuscript
considerably.
We acknowledge the use of {\it Hinode}/EIS observations.
AKS acknowledges the support of the International Exchange Scheme (Royal Society) between Indian and UK.
The software used in this work was in part developed by the DOE-supported ASCI/Alliance Center 
for Astrophysical Thermonuclear Flashes at the University of Chicago. 
\end{acks}



\end{document}